\documentclass[12pt]{article}
\usepackage{amsmath}
\usepackage{graphicx}
\usepackage{enumerate}
\usepackage{natbib} 
\setcitestyle{aysep={}}    
\setcitestyle{citesep={;}} 
\usepackage{url} 
\usepackage[title]{appendix}
\RequirePackage[hidelinks]{hyperref}

\RequirePackage[OT1]{fontenc}
\RequirePackage{amsthm,amsmath}
\RequirePackage{natbib}
\RequirePackage{graphicx}

\usepackage{bbm}
\usepackage{algorithm2e}
\usepackage{enumitem}
\usepackage{multirow}
\usepackage{subcaption}

\usepackage{amsmath,amsfonts,amssymb}
\usepackage{multirow}
\usepackage{graphicx}
\usepackage{rotating,tabularx}

\usepackage[utf8]{inputenc}
\usepackage[english]{babel}
\usepackage{amsthm}
\usepackage{booktabs}
\usepackage{algorithmic}

\newcommand{\blind}{0}

\newcommand{\R}{\mathbb{R}} 
\newcommand{\Rn}{\mathbb{R}^n} 
\newcommand{\Rp}{\mathbb{R}^p} 
\newcommand{\Rnp}{\mathbb{R}^{n\times p}} 
\newcommand{\ra}[1]{\renewcommand{\arraystretch}{#1}}

\newcommand{\E}{\mathbb{E}} 

\DeclareMathOperator*{\argmin}{arg\,min}

\newcommand{\samplesize}{n} 
\newcommand{\numvar}{p} 
\newcommand{\samx}{\boldsymbol{x}}
\newcommand{\samxi}{\boldsymbol{{x}}_i} 
\newcommand{\variance}{\ensuremath{\sigma}}
\newcommand{\distribution}{\ensuremath{F}}

\newcommand{\data}{\boldsymbol{Z}}
\newcommand{\partiald}{\boldsymbol{\partial}}

\newcommand{\derivative}{\ensuremath{\boldsymbol{s}}}
\newcommand{\samyi}{y_i} 
\newcommand{\samy}{y}
\newcommand{\design}{\boldsymbol{X}} 
\newcommand{\outcome}{\boldsymbol{y}} 
\newcommand{\vare}{\boldsymbol{\varepsilon}}
\newcommand{\varei}{\varepsilon_i}
\newcommand{\vares}{\varepsilon}
\newcommand{\zero}{\ensuremath{\boldsymbol{0}}}

\newcommand{\target}{\boldsymbol{\beta}^*} 
\newcommand{\parameter}{\boldsymbol{\beta}} 
\newcommand{\est}{\widehat{\parameter}} 
\newcommand{\paratheta}{\boldsymbol{\theta}} 
\newcommand{\esttheta}{\boldsymbol{\widehat{\theta}}}
\newcommand{\targettheta}{\boldsymbol{\theta^{*}}}
\newcommand{\paraspace}{\Rp}
\newcommand{\paraphi}{\phi} 
\newcommand{\tuning}{\lambda} 
\newcommand{\ridgetuning}{\tau} 
\newcommand{\margincondC}{C} 
\newcommand{\esttuning}{\widehat{\tuning}} 

\newcommand{\linkfcnn}{g} 

\newcommand{\fcnnb}{b}
\newcommand{\fcnnc}{c}
\newcommand{\fcnnf}{f}

\newcommand{\fcnb}[1]{\ensuremath{b({#1})}}
\newcommand{\fcnc}[1]{\ensuremath{c({#1})}}
\newcommand{\objfcnTe}{D} 

\newcommand{\tp}{^{\top}} 
\newcommand{\inv}{^{-1}} 
\newcommand{\sumin}{\sum_{i=1}^{n}} 

\newcommand{\normone}[1]{\|{#1}\|_1} 
\newcommand{\normtwo}[1]{\ensuremath{|\!|#1|\!|_{2}}} 
\newcommand{\normtwos}[1]{\|{#1}\|_2^2} 
\newcommand{\norminfty}[1]{\|{#1}\|_\infty} 

\newcommand{\bc}[1]{\ensuremath{\{{#1}\}}} 
\newcommand{\bcbb}[1]{\ensuremath{\Big\{{#1}\Big\}}} 
\newcommand{\bcbbb}[1]{\ensuremath{\bigg\{{#1}\bigg\}}} 
\newcommand{\bcbbbb}[1]{\ensuremath{\Bigg\{{#1}\Bigg\}}} 

\newcommand{\prt}[1]{\ensuremath{({#1})}} 
\newcommand{\prtb}[1]{\ensuremath{\big({#1}\big)}} 
\newcommand{\prtbb}[1]{\ensuremath{\Big({#1}\Big)}} 
\newcommand{\prtbbb}[1]{\ensuremath{\bigg({#1}\bigg)}} 

\newcommand{\loglikelihood}{l}

\newcommand{\N}{\mathcal{N}} 

\newcommand{\nameedr}{\ensuremath{\operatorname{edr}}}

\newcommand{\nametrex}{\ensuremath{\operatorname{trex}}}

\newcommand{\nameridge}{\ensuremath{\operatorname{ridge}}}
\newcommand{\namemethod}{\ensuremath{\operatorname{t-ridge}}}
\newcommand{\subdifferential}{\ensuremath{\operatorname{\boldsymbol{\kappa}}}}
\numberwithin{equation}{section}
\newtheorem{thm}{Theorem}[section]
\newtheorem{lemma}{Lemma}[section]

\newtheorem{condition}{Condition}[section]

\usepackage{xcolor}

\begin{document}

\def\spacingset#1{\renewcommand{\baselinestretch}%
{#1}\small\normalsize} \spacingset{1}


\if0\blind
{
  \title{\bf Tuning-free ridge estimators for high-dimensional generalized linear models}
\author{Shih-Ting Huang, Fang Xie, and Johannes Lederer 
 \\
 Ruhr-Universit\"at Bochum, 44801 Bochum, Germany}
 \maketitle
} \fi

\if1\blind
{
  \bigskip
  \bigskip
  \bigskip
  \begin{center}
    {\LARGE\bf Tuning-free ridge estimators for high-dimensional generalized linear models}
\end{center}
  \medskip
} \fi

\bigskip
\begin{abstract}
Ridge estimators regularize the squared Euclidean lengths of parameters.
Such estimators are mathematically and computationally attractive but involve tuning parameters that can be difficult to calibrate.
In this paper, we show that ridge estimators can be modified such that tuning parameters can be avoided altogether. 
We also show that these modified versions can improve on the empirical prediction accuracies of standard ridge estimators combined with cross-validation, and we provide first theoretical guarantees.
\end{abstract}

\noindent%
\vfill

\newpage
\spacingset{1.5} 

\section{Introduction}
\label{sec:introduction}

High-dimensional estimators typically minimize an objective function that contains again two functions:
a data-fitting function to ensure a good fit to the data and a penalty function to leverage additional information.
Popular data-fitting functions are least-squares and negative log-likelihood;
popular penalty functions are $\ell_{1}$ (lasso)~\citep{tibshirani_1996} and $\ell_{2}^2$ (ridge)~\citep{hoerl1970ridge}.
The weighting between data-fitting and penalty function is finally determined by a tuning parameter,
which needs to be calibrated to fit the specific estimator, data, and application at hand.

Known calibration schemes such as cross-validation~\citep{stone1974cross, golub1979generalized}, 
stability selection~\citep{meinshausen_buhlmann_2010,Shah2013Variable},
and adaptive validation~\citep{Chichignoud2016, li_lederer_2019, Mahsa2019} require two steps:
compute the estimators or surrogates of them for a range of tuning parameters and then apply a rule to select among those candidate estimators.
We now focus on the ridge estimators and pose the question of whether the calibration of their tuning parameters can instead be integrated into the estimation process directly.

In this paper,
we modify standard ridge estimators such that the calibration of the tuning parameter is indeed part of the estimation process directly.
We make use of two earlier lines of research: 
First, the edr~\citep{STHuang2019},
which shows 
that replacing $\ell_2^2$-regularization by $\ell_2$-regularization can make estimators amenable to recent techniques in high-dimensional theory.
Second,
the trex~\citep{LedererandMuller2015, bien_gaynanova_lederer_muller_2018, bien-gaynanova-lederer-muller-2018, LedererandMuller2014},
which proposes a way to integrate tuning parameter calibration into lasso-type estimators.
However, while both of these lines of research focus on regularized least-squares in linear regression,
we demonstrate that an
inherent calibration of the ridge parameter is possible for a wide range of data-fitting functions and models.

We make three main contributions:
\begin{itemize}
    \item We motivate  alternative ridge estimators that dispense with tuning parameters (Section~\ref{sec:methodology}).
    \item We establish theoretical insights for these new estimators (Theorems~\ref{thm:edr-tridge}, \ref{thm:uniqueness},  and~\ref{thm:t-ridge}) and also for the underlying edr estimators (Theorems \ref{thm:ridge-edr} and~\ref{thm:edrBound}).
    \item We show that the tuning-free estimators can be readily computed (Section~\ref{subsec:algorithm}) and rival our outmatch standard pipelines empirically (Section~\ref{subsec:simulation}).
\end{itemize}


\section{Methodology}
\label{sec:methodology}

Standard methods to estimate a target parameter $\targettheta \in \Rp$ from data~$\data$ are  ridge-type estimators of the form
\begin{equation}
\label{eq:generalizedRidge}
\esttheta_{\nameridge}[\ridgetuning] \in \argmin\limits_{\paratheta \in \paraspace}\bcbb{\objfcnTe\prt{\paratheta|\data}+\ridgetuning\normtwo{\paratheta}^2}, 
\end{equation}
\noindent where $\objfcnTe\prt{\paratheta|\data}:\Rp\to\R$ is a data-fitting function and $\ridgetuning\in[0,\infty)$ is a tuning parameter.
Ridge regularization, also known as Tikhonov regularization, can be traced back to~\citep{Tikhonov1943ridge}. 
A common data-fitting function is the least-squares function $    \objfcnTe\prt{\paratheta|\data}:=\normtwo{\outcome-\design\paratheta}^2$ for regression  data $\data=(\outcome, \design)\in\R^n\times \Rnp$,
which leads to the usual ridge estimator~\citep{hoerl1970ridge}.
Well-known extensions of this estimator define  $\objfcnTe\prt{\paratheta|\data}$ as negative log-likelihood functions~\citep{glm}.

A main challenge in the application of these estimators is the calibration of~$\ridgetuning$.
Our objective is, therefore, to rewrite the estimators such that we can avoid this tuning parameter.
Our first step is to change
the $\ell_{2}^2$-prior function in~\eqref{eq:generalizedRidge} to~$\ell_{2}$:
\begin{equation}
    \label{eq:generalizedEDR}
\esttheta_{\nameedr}[\tuning] \in \argmin\limits_{\paratheta\in\paraspace}\bcbb{\objfcnTe\prt{\paratheta | \data}+\tuning\normtwo{\paratheta}}.
\end{equation}
These estimators generalize the edr estimator for linear regression~\citep{STHuang2019}. 
We will see in the following 
that the change from~\eqref{eq:generalizedRidge} to~\eqref{eq:generalizedEDR} allows us to apply standard techniques from modern high-dimensional theory while preserving the original estimators' key features such as their computational simplicity.

Indeed, edr and ridge estimators are computational siblings. 
Assuming---for simplicity---here and in the following that the data-fitting function~$\objfcnTe$ is convex and differentiable, we can define the ``score" function as 
\begin{equation}
\label{eq:score}
\derivative\prt{\paratheta} :=
-\frac{\partiald\objfcnTe\prt{\paratheta|\data}}{\partiald\paratheta}
\end{equation}  
\noindent and find the following (all proofs are deferred to  Appendix~\ref{subsec:proofs}):
\begin{thm}[Equivalence of edr and ridge]
\label{thm:ridge-edr}
Edr estimator $\esttheta_{\nameedr}[\tuning]$ and ridge estimator $\esttheta_{\nameridge}[\ridgetuning]$ are equivalent if the following two statements hold: 
\begin{enumerate}
    \item For each ridge estimator $\esttheta_{\nameridge}[\ridgetuning]$ with $\ridgetuning\geq 0$, there exists a $\tuning=2\ridgetuning\normtwo{\esttheta_{\nameridge}[\ridgetuning]}\geq0$ such that $\esttheta_{\nameedr}[\tuning]=\esttheta_{\nameridge}[\ridgetuning]$;
    \item For each edr estimator $\esttheta_{\nameedr}[\tuning]$ with $\tuning \geq 0$, there exists a $\ridgetuning=\tuning / (2\normtwo{\esttheta_{\nameedr}[\tuning]})\geq0$ such that $\esttheta_{\nameridge}[\ridgetuning]=\esttheta_{\nameedr}[\tuning]$.
    In particular, if $\esttheta_{\nameedr}[0]=\zero_{\numvar}$, then there exists $\ridgetuning= 0$ such that  $\esttheta_{\nameridge}[0]=\esttheta_{\nameedr}[0]=\zero_{\numvar}$.
\end{enumerate}
Moreover, if $\esttheta_{\nameedr}[\tuning]=\esttheta_{\nameridge}[\ridgetuning]$, then
\begin{equation*}
    \tuning= \normtwo{\derivative\prt{\esttheta_{\nameedr}[\tuning]}}=\normtwo{\derivative\prt{\esttheta_{\nameridge}[\ridgetuning]}}.
\end{equation*}

\end{thm}
\noindent This result generalizes Theorem~2 of~\cite{STHuang2019} for the edr estimator in linear regression (that special case also follows from~\citep[Theorem~5]{ahsen_vidyasagar_2017} for the clot estimator,
which combines $\ell_1$- and $\ell_2$-regularization.).  
It~shows first that the ridge and edr paths are remappings of each other and~then gives a relationship between the edr tuning parameter and the score function.
These two observations are crucial for the following.

Our second step is to modify the data-fitting function $\objfcnTe\prt{\paratheta | \data}$ in  \eqref{eq:generalizedEDR} in a way that makes tuning parameters unnecessary.
Our motivation comes from the trex in $\ell_1$-regularized linear regression~\citep{LedererandMuller2015}:
\begin{equation*}
\esttheta_{\nametrex} \in \argmin\limits_{\paratheta \in \paraspace}\bcbbbb{\frac{\normtwo{\outcome-\design\paratheta}^2}{\norminfty{\design\tp\prt{\outcome-\design\paratheta}}/2}+\normone{\paratheta}}.
\end{equation*}
The idea of the trex is to amend the lasso estimator~\citep{tibshirani_1996}   with the additional factor $\norminfty{\design\tp\prt{\outcome-\design\paratheta}}/2$  for an ``inherent" calibration of the tuning parameter.
This modification is unsuitable for us because our estimators include general data-fitting functions and a different regularizer,
but we can still use that overall idea of complementing the data-fitting function with a factor.
The factor is motivated by Theorem~\ref{thm:ridge-edr}: 
we divide the objective function in~\eqref{eq:generalizedEDR} by~$\tuning$ and then replace~$\tuning$ by its ``functional value"  $\tuning=\normtwo{\derivative\prt{\esttheta_{\nameedr}[\tuning]}}$.
We call the resulting estimator 
\begin{equation}
    \label{eq:generalizedt-ridge}
\esttheta_{\namemethod} \in \argmin\limits_{\paratheta \in \paraspace}\bcbbbb{\frac{\objfcnTe\prt{\paratheta|\data}}{\normtwo{\derivative\prt{\paratheta}}}+\normtwo{\paratheta}}
\end{equation}
the \emph{t-ridge}.
Similarly as the trex,
the t-ridge does away with tuning parameters. 

In contrast to trex that requires elaborate algorithms~\citep{bien_gaynanova_lederer_muller_2018}, we first show that the t-ridge is simply one element of the path of the ridge estimator. 

\begin{thm}[T-ridge is on the ridge path]
\label{thm:edr-tridge}
Let $\esttheta_{\namemethod}$ be a t-ridge estimator defined by~\eqref{eq:generalizedt-ridge}. 
Define $\tuning:=\normtwo{\derivative\prt{\esttheta_{\namemethod}}}$. 
Then, there always exists a tuning parameter $\ridgetuning := \tuning/(2\normtwo{\esttheta_{\nameedr}[\tuning]})$ such that 
$\esttheta_{\nameridge}[\ridgetuning]$ minimizes the objective function in~\eqref{eq:generalizedt-ridge} and
$\esttheta_{\namemethod}=\esttheta_{\nameedr}[\tuning]=\esttheta_{\nameridge}[\ridgetuning]$. 
\end{thm}

Theorem~\ref{thm:edr-tridge} also implies that t-ridge is on the path of edr. 
Moreover, under a mild technical assumption, the t-ridge estimator is unique.
\begin{thm}[Uniqueness of the t-ridge estimator]
\label{thm:uniqueness}
If $\objfcnTe\prt{\paratheta|\data}>0$ for any $\paratheta \in \Rp$, the minimum of the t-ridge objective function in~\eqref{eq:generalizedt-ridge} is unique.
\end{thm}
\noindent Such a result has not been established for the trex estimator. 
It ensures that the t-ridge estimator retains the uniqueness of the ridge estimator. 
Hence, calculating the t-ridge essentially amounts to a grid search on the ridge path--see Section~\ref{subsec:algorithm} for details.

\section{Applications in generalized linear models}
\label{sec:applications}
We now apply the t-ridge estimator to generalized linear models and derive the first theoretical results.

\subsection{T-ridge estimator for generalized linear models}
In this section, we exemplify the t-ridge estimator for maximum regularized likelihood estimation in generalized linear models. 
We consider data $\data=(\outcome, \design)$ that follow a conditional distribution 
\begin{equation}
\label{eq:model}
\samyi|\samxi, \target \sim \distribution\ \ \ {\rm with}\ \linkfcnn\prtb{\E\prt{\samyi|\samxi, \target}}=\samxi\tp\target.
\end{equation}
\noindent Here, $\outcome=(\samy_1,\ldots,\samy_n) \in \Rn$ is a vector of outcomes and $\design=(\samx_1,\ldots,\samx_n)\tp \in \Rnp$ a design matrix.
The distribution~$\distribution$ is assumed in the exponential family, $\linkfcnn:\R\mapsto\R$ is a link function, and $\target\in\Rp$ is the unknown regression vector.
We allow for high-dimensional settings, that is, the number of parameters~$\numvar$ may rival or even exceed the number of observations~$\samplesize$.

For every vector $\parameter \in \Rp$, 
the density of $\samyi|\samxi, \parameter$ can be written as~\citep{glm}
\begin{equation*}
\fcnnf\prt{\samyi|\samxi, \parameter}=\exp\bcbbb{\frac{\samyi\samxi\tp\parameter-\fcnb{\samxi\tp\parameter}}{d(\paraphi)}+\fcnc{\samyi,\paraphi}}.
\end{equation*}
The inner product $\samxi\tp\parameter$ is related to the mean of the distribution and the dispersion parameter~$\paraphi \in \R$ to the variance of the distribution;
indeed,
the mean of $\samyi|\samxi, \parameter$ is $\linkfcnn\inv\prt{\samxi\tp\parameter}$, and the variance of $\samyi|\samxi, \parameter$ is $d(\paraphi)$ times the second derivative of the function $\fcnnb$ with respect to $\samxi\tp\parameter$.
Without loss of generality, each outcome $\samyi, i \in \{1, \dots, \samplesize\}$ can be written as its' true mean $\E\prt{\samyi|\samxi, \target}=\linkfcnn\inv\prt{\samxi\tp\target}$ plus a random noise. That is, 
\begin{equation*}
    \samyi = \linkfcnn\inv\prt{\samxi\tp\target} + \varei,\ \ \ {\rm for}\ i\in \bc{1,\ldots,\samplesize}, 
\end{equation*}
\noindent where $\varei \in \R$ are the random noises.
The real-valued functions~$\fcnnb$, $\fcnnc$, $d$, and~$\linkfcnn\inv$ are specified by the concrete choice of  the distribution~$\distribution$; their forms for the most common distributions are given in Table~\ref{table:glm}. 
To be clear, we consider the canonical link function, which satisfies $\linkfcnn\inv\prt{z}=\fcnnb'\prt{z}$ for any $z\in\R$.

Assuming that the $\samyi$'s are independent, the log-likelihood function of $\distribution$ is
\begin{align*}
\label{eq:loglikelihod}
    \loglikelihood\prt{\parameter|\outcome, \design}&=\log\prtbbb{\prod_{i=1}^{n}\fcnnf\prt{\samyi|\samxi\tp\parameter,\paraphi}} \\
    &=\sumin\prtbbb{\frac{\samyi\samxi\tp\parameter-\fcnb{\samxi\tp\parameter}}{d(\paraphi)}+\fcnc{\samyi,\paraphi}}.
\end{align*}
Omitting factors and summands that do not depend on~$\parameter$,
we find the negative log-likelihood data-fitting term can be simplified as 
\begin{equation}
\label{eq:glmlog}
   \objfcnTe\prtb{\parameter|\prt{\outcome, \design}} = -\sumin\prtb{\samyi\samxi\tp\parameter-\fcnb{\samxi\tp\parameter}}.
\end{equation}

\begin{table}
\caption{common distributions with their $\fcnnb$, $\fcnnc$, $d$, and $\linkfcnn\inv$ functions}
\centering
\ra{1.5}
\begin{tabular}{ccccc}
\toprule
\midrule
distribution of $\distribution$ & $\fcnb{\samxi\tp\parameter}$ &  $\linkfcnn\inv\prt{\samxi\tp\parameter}$ & $\fcnc{\samyi,\paraphi}$
& $d(\paraphi)$\\
\midrule
Gaussian & $\dfrac{\prt{\samxi\tp\parameter}^{2}}{2}$ & $\samxi\tp\parameter$ & $\dfrac{1}{2d(\paraphi)}\samyi^{2}-\dfrac{1}{2}\log\prtb{2\pi d(\paraphi)}$
& $\sigma^{2}$ \\
Poisson & $\exp\prt{\samxi\tp\parameter}$ & $\exp\prt{\samxi\tp\parameter}$ & $-\log(\samyi!)$
& $1$\\
Bernoulli & $\log\prt{1 + \exp\prt{\samxi\tp\parameter}}$ & $\dfrac{\exp\prt{\samxi\tp\parameter}}{1 + \exp\prt{\samxi\tp\parameter}}$ & 
0 
& $1$\\
\midrule
\bottomrule
\end{tabular}
\label{table:glm}
\end{table}

Observing that the derivative of function $\fcnnb$ is $\linkfcnn\inv$,
the corresponding score function~\eqref{eq:score} is
\begin{equation}
\label{eq:scoreglm}
    \derivative\prt{\parameter} =-\sumin\samxi\tp\prtb{\samyi-\linkfcnn\inv\prt{\samxi\tp\parameter}}.
\end{equation}
In view of~\eqref{eq:generalizedt-ridge},
this means that the  t-ridge estimator is 
\begin{equation}
\label{eq:t-ridgeglm}
 \est_{\namemethod}\in\argmin\limits_{\parameter\in\Rp}\bcbbbb{\frac{-\sumin\prtb{\samyi\samxi\tp\parameter-\fcnb{\samxi\tp\parameter}}}{\normtwo{\sumin\samxi\tp\prtb{\samyi-\linkfcnn\inv\prt{\samxi\tp\parameter}}}}+\normtwo{\parameter}}.
\end{equation}
\noindent As compared to the general form of the t-ridge in~\eqref{eq:generalizedt-ridge}, the general parameter~$\paratheta$ is specified to the regression vector~$\parameter$, the general data~$\data$ is specified to the regression data $(\outcome, \design)$,
and the data-fitting function~$\objfcnTe$ and the score function $\derivative$ are specified to \eqref{eq:glmlog} and \eqref{eq:scoreglm}, respectively.

\subsection{Further theoretical insights}

We now establish theoretical insights into the t-ridge estimator. 
Our two results are 
Theorem~\ref{thm:edrBound},
which is a novel prediction guarantee for (a generalized version of) the related edr estimator, and Theorem~\ref{thm:t-ridge},
which is a prediction guarantee for the t-ridge estimator.



We first introduce a standard ``margin condition'' for the function $\fcnnb$. 
\begin{condition}[Margin condition for the function $\fcnnb$]
\label{cond:margin}
For each $\samxi, i \in \{1, \dots, \samplesize\}$,  if a given vector $\parameter \in \Rp$ such that $\normtwo{\parameter - \target} \leq \delta$ for some constants $\delta > 0$, then there exist a constant $\margincondC_{i} > 0$ with
\begin{equation*}
    \fcnb{\samxi\tp\parameter} - \fcnb{\samxi\tp\target}\ge \fcnnb'(\samxi\tp\target)(\samxi\tp\parameter-\samxi\tp\target)+\frac{1}{\margincondC_{i}^{2}}(\samxi\tp\parameter-\samxi\tp\target)^2.
\end{equation*}


\end{condition}
\noindent Notice that the function $\fcnnb$ for generalized linear model is differentiable, so $\fcnnb'(\cdot)$ always exists. 
Such a constant~$\margincondC_{i}$ always exists in generalized linear models~\citep[Section~11.6]{Van2016Estimation} once $\samxi, \parameter \in \Rp$ are given;
for example, $\margincondC_{i}=2, i \in \{1, \cdots, \samplesize\}$ in the case that $\distribution$ satisfies the Gaussian distribution. 
The existence of the constants implies in particular that the function~$\fcnnb$ is strictly convex.

The following theorem shows that $\tuning^*:=\normtwo{\derivative\prt{\target}}=\normtwo{\design\tp\vare}$, 
where $\vare=\prt{\vares_1,\ldots,\vares_n}\tp$ is the noise vector, is indeed an optimal tuning parameter for the edr estimator.
This result further strengthens our motivation for the t-ridge estimator 
as the edr estimator calibrated inherently to that~$\tuning$.

\begin{thm}[Prediction error bound for the generalized edr]
\label{thm:edrBound}
Consider data $\prt{\outcome, \design}$ that follow~\eqref{eq:model} and the corresponding edr estimator 
\begin{equation}
\label{eq:glmEdr}
    \est_{\nameedr}[\tuning]\in\argmin\limits_{\parameter\in\Rp}\bcbbbb{-\sumin\prtb{\samyi\samxi\tp\parameter-\fcnb{\samxi\tp\parameter}}+\tuning\normtwo{\parameter}}
\end{equation}
\noindent according to~\eqref{eq:generalizedEDR}. 
With~$\margincondC=max_{i \in \{1, \cdots, \samplesize\}}\{C_{i}\}$\ as in the  margin condition~\ref{cond:margin} and define $\tuning^{*}:=\normtwo{\derivative\prt{\target}}$,
it holds for $\tuning\ge\tuning^*$ that 
\begin{equation*}
    \normtwos{ \design\prt{\est_{\nameedr}[\tuning]-\target}}\le 2\margincondC^2\tuning\normtwo{\target}.
\end{equation*}
\end{thm}
\noindent
This bound complements the  bound on the individual prediction errors~$|\samxi\tp\prt{\est_{\nameedr}[\tuning]-\target}|$ that has been derived previously~\citep[Lemma~1]{STHuang2019}. 
But more interesting here is that it provides further support for the t-ridge estimator:
The bound suggests that for accurate prediction with the edr estimator,
the tuning parameter~$\tuning$ needs to be sufficiently small (since the bound is proportional to $\tuning$), but not too small (to satisfy the condition $\tuning \geq \normtwo{\design\tp\vare}$). 
In particular,
the bound is optimized at $\tuning=\normtwo{\derivative\prt{\target}}=\normtwo{\design\tp\vare}$---in line with our motivation for the t-ridge estimator. 

The following theorem finally gives a  bound on the prediction loss of the t-ridge estimator.

\begin{thm}[Prediction error bound for the t-ridge]
\label{thm:t-ridge}
Consider data $\prt{\outcome, \design}$ that follow~\eqref{eq:model} and t-ridge estimator $\est_{\namemethod}$ defined in~\eqref{eq:t-ridgeglm}. 
Let $\esttuning:=\normtwo{\derivative\prt{\est_{\namemethod}}}$ and $\tuning^{*}:=\normtwo{\derivative\prt{\target}}$. 
With~$\margincondC=max_{i \in \{1, \cdots, \samplesize\}}\{C_{i}\}$\ as in the  margin condition~\ref{cond:margin},
it holds that
\begin{equation*}
    \normtwos{\design\prt{\est_{\namemethod}-\target}}\le 2\margincondC^2\max\{\tuning^{*},\esttuning\}\normtwo{\target}.
\end{equation*}

\end{thm}
\noindent 
This bound parallels the one for the trex for $\ell_1$-regularized linear regression~\citep[Theorem~2]{bien-gaynanova-lederer-muller-2018}. 
Moreover, it relates to Theorem~\ref{thm:edrBound};
in particular, if $\esttuning\le\tuning^*$,
then the t-ridge bound equals the edr bound at the \emph{optimal} tuning parameter~$\tuning^*$---\emph{without}  the t-ridge knowing that tuning parameter.
Interestingly, one can check if this case applies  in an extremely simple way:
\begin{lemma}[Relationship between $\tuning^{*}$ and $\esttuning$]
\label{lem:checktridge}
If $\objfcnTe\prt{\est_{\namemethod}|\outcome, \design} > 0$, then $\esttuning \ge \tuning^{*}$; 
if $\objfcnTe\prt{\est_{\namemethod}|\outcome, \design} < 0$, then $\esttuning \le \tuning^{*}$. 
\end{lemma}
\noindent Together with Theorem~\ref{thm:t-ridge},
this gives a concrete guarantee for the prediction accuracy of the t-ridge.


\section{Algorithm and Numerical Analysis}
\label{sec:numerical}

In this section, 
we introduce a specific algorithm for the  t-ridge estimator  for maximum regularized likelihood estimation in generalized linear models. 
We then show that the t-ridge  matches or even outperforms ridge combined with cross-validation,  the standard pipeline in this context, in the three most common cases for
the distribution~$\distribution$: Gaussian, Poisson, and Bernoulli. 

\subsection{Algorithm}
\label{subsec:algorithm}

The t-ridge's objective function for maximum regularized likelihood estimation in generalized linear regression 
\begin{equation}
\label{eq:objectglm}
    \fcnnf_{\namemethod}: \parameter \rightarrow
    \frac{-\sumin\prtb{\samyi\samxi\tp\parameter-\fcnb{\samxi\tp\parameter}}}{\normtwo{\sumin\samxi\prt{\samyi-\linkfcnn\inv\prt{\samxi\tp\parameter}}}}+\normtwo{\parameter}
\end{equation}
seems very hard to optimize,
in particular,
because it is non-convex.
But Theorem~\ref{thm:edr-tridge} entails a very simple and effective optimization strategy:
solve the ridge path with a standard algorithm  and then select the solution that minimizes the t-ridge objective function~\eqref{eq:objectglm}. 

It turns out that this strategy can be improved  even further:
one can use the differentiability of the objective function~\eqref{eq:objectglm} to speed up the grid search over the ridge solution path.
We proceed in three steps:
\begin{enumerate}[leftmargin=1.8cm]
    \item[\textit{Step 1:}] Compute a stationary point $\est_{\operatorname{sp}}$ of the t-ridge object function~\eqref{eq:objectglm} with a standard algorithm such as the Fletcher-Reeves algorithm~\citep{fletcher_1964}.
    
    \item[\textit{Step 2:}] Compute the ridge tuning parameter $\ridgetuning:=\normtwo{\derivative\prt{\est_{\operatorname{sp}}}} / (2\normtwo{\est_{\operatorname{sp}}})$ and set $\ridgetuning_{\operatorname{min}}:=\max\{0.05,(\ridgetuning-c)\}$ and   $\ridgetuning_{\operatorname{max}}:=\ridgetuning+c$ for a given range $c\in(0,\infty)$.  
    
    \item[\textit{Step 3:}] Compute the ridge estimator with a standard algorithm for $m$ equally spaced tuning parameters in  $[\ridgetuning_{\operatorname{min}}, \ridgetuning_{\operatorname{max}}]$, and then select the corresponding estimator that minimizes the t-ridge object function~\eqref{eq:objectglm}.
\end{enumerate}
The underpinning idea is that the stationary points of~\eqref{eq:objectglm} give a hint of what ridge estimators are relevant so that the search over the ridge path can be narrowed down to a small interval.
And we indeed find empirically that the gain that is due to restricting the ridge  path (Step~3) outweighs the computations of the stationary point (Step~1) and the two ridge estimators (Step~2). 

Another statement of our approach is Algorithm~\ref{alg:tridge}.

\begin{algorithm}[htb]
   \caption{t-ridge in generalized linear models}
   \label{alg:tridge}
\begin{algorithmic}
   \STATE {\bfseries Input:} data $\prt{\outcome, \design}$, range $c$, and number of candidate ridge tuning parameters $m$ \\
   {\bfseries Output:} $\est_{\namemethod}$ from~\eqref{eq:t-ridgeglm}
      \STATE
   Compute a stationary point $\est_{\operatorname{sp}}$ of~\eqref{eq:objectglm} using  the Fletcher-Reeves algorithm\\
   \STATE 
   Set $\ridgetuning
   :=\normtwo{\derivative\prt{\est_{\operatorname{sp}}}} / (2 \normtwo{\est_{\operatorname{sp}}})$ 
   \IF{$\est_{\operatorname{sp}} = \zero_{\numvar}$}
        \STATE $\ridgetuning_{\operatorname{min}}:=10^{10}
        ;
        \ridgetuning_{\operatorname{max}}:=10^{11}
        $ \\
   \ELSE 
        \STATE $\ridgetuning_{\operatorname{min}}:= \max\{0.05,\ridgetuning-c\}$ \\
        \STATE $\ridgetuning_{\operatorname{max}}:= \ridgetuning + c$ \\
   \ENDIF \\
   
   \FOR{$i=1$ {\bfseries to} $m$
   }
   \STATE 
   $\ridgetuning_{i} := \ridgetuning_{\operatorname{min}} + (\ridgetuning_{\operatorname{max}}-\ridgetuning_{\operatorname{min}}) \cdot i / m$ \\
   Compute $\est_{\nameridge}[\ridgetuning_{i}]$ and  $\fcnnf_{\namemethod}\prt{\est_{\nameridge}[\ridgetuning_{i}]}$ 
   \ENDFOR 
   \STATE
   Set $\hat{\ridgetuning} \in \argmin\limits_{i\in\{1,\dots,m\}}\{\fcnnf_{\namemethod}\prt{\est_{\nameridge}[\ridgetuning_{i}]}\}$ 
   \STATE {\bfseries Return} $\est_{\namemethod}=\est_{\nameridge}[\hat{\ridgetuning}]$
   
\end{algorithmic}
\end{algorithm}

Throughout, 
we set $c:=0.1$ and  $m:=1000$,
which leads to excellent results over a wide range of settings.
As a technical detail, 
we set $\ridgetuning_{\min}:=10^{10}$ and $\ridgetuning_{\max}:=10^{11}$
if
$\normtwo{\est_{\operatorname{sp}}}=0$ to avoid vanishing denominators,
and for similar reasons, we set $\ridgetuning_{\operatorname{min}}:=\max\{0.05,\ridgetuning-c\}$.


\subsection{Numerical Analysis}
\label{subsec:simulation}

We now show that our pipeline rivals $K$-fold cross-validation,
the standard pipeline in this context.
We compute the latter with the \texttt{glmnet} package in~\texttt{R} with default settings~\citep{glmnet}.

The dimensions of the design matrix are $(n,p) \in \{ (100,300), (200,500), (50,1000) \}$. 
Each row $\samxi \in \Rp$  
of the design matrix $\design \in \Rnp$ is sampled from a $p$-dimensional normal distribution with mean~$\mathbf{0}_p$ and covariance matrix $\boldsymbol{\Sigma}$, where ${\Sigma}_{uv}=k^{|u-v|}$, $u,v\in\bc{1, \dots, \numvar}$, and $k \in \{0, 0.2, 0.4 \}$ is the magnitude of the mutual correlations
($0^{0}:=1$). 
The columns of the design matrix are then normalized to have Euclidean norm equal to one. 
The entries of the regression vector~$\target$ are sampled i.i.d.\@ from $\N\prt{0,1}$ and then projected onto the row space of~$\design$ to ensure identifiability~\citep{Shao2012, Buhlmann2013}.

We run 100 experiments for each set of parameters and report the means of the relative prediction 
errors defined by
$\normtwo{\design\prt{\est-\target}}/\normtwo{\design\target}$. 

\subsubsection{Gaussian case}
\label{subsec:examplelgaussian}

We first generate Gaussian data, where the outcome vector
\begin{equation*}
    \outcome = \design\target + \vare
\end{equation*}
\noindent is the true signal $\design\target$ plus the noise vector $\vare$.
The entries of the noise vector~$\vare$ are sampled i.i.d.\@ from $\N\prt{0,\variance^2}$, where $\variance^2$ is set such that the  signal-to-noise ratio 
\begin{equation*}
    \frac{\prtbb{\sumin\prt{\samxi\tp\target}^{2}-\frac{\prtb{\sumin\samxi\tp\target}^{2}}{\samplesize}}}{\variance^{2}\prt{\samplesize-1}}
\end{equation*}
\noindent equals~10.                             
According to Table~\ref{table:glm},
the t-ridge  estimator~\eqref{eq:t-ridgeglm} is
\begin{equation*}
    \est_{\namemethod} \in \argmin\limits_{\parameter\in\Rp}\bcbbbb{\frac{\normtwo{\outcome-\design\parameter}^{2}- \normtwo{\outcome}^2}{2\normtwo{\design\tp\prt{\outcome-\design\parameter}}} +\normtwo{\parameter}}.
\end{equation*}  

\begin{table}[htb]
    \centering
    \caption{t-ridge outperforms $K$-fold cross-validated ridge ($K$-fold CV ridge) for $K \in \{5, 10\}$  in prediction on  Gaussian data with $k=0$.}
    \ra{1.2}
    \begin{tabular}{c c c c c c}
    \toprule
    \midrule
         \multirow{2}{*}{Relative prediction error} &
         \multirow{2}{*}{n} & 
         \multirow{2}{*}{p} & 
         \multicolumn{3}{c}{Mean of relative errors (sd)} \\ \cline{4-6}
         
         {} & {} & {} & t-ridge 
                      & 5-fold CV ridge
                      & 10-fold CV ridge \\
         \hline
         \multirow{3}{*}{$\frac{\normtwo{\design\prt{\est-\target}}}{\normtwo{\design\target}}$} &
              100 & 300 & 0.34 (0.03) & 0.53 (0.05) & 0.53 (0.04) \\ 
{} & 200 & 500 & 0.36 (0.02) & 0.51 (0.02) & 0.51 (0.02) \\ 
{} & 50 & 1000 & 0.31 (0.04) & 0.57 (0.28) & 0.54 (0.28) \\  
         \midrule
         \bottomrule
    \end{tabular}
    \label{table:gaussian0}
\end{table}

Table~\ref{table:gaussian0} demonstrates that the t-ridge outperforms  5- and 10-fold  cross-validated ridge. 
(The results for $k \in \{0.2, 0.4\}$ are deferred to Table~\ref{table:gaussian} in Appendix~\ref{sec:addition}).

\begin{figure}[htb]
    \centering
    \includegraphics[width=0.7\textwidth]{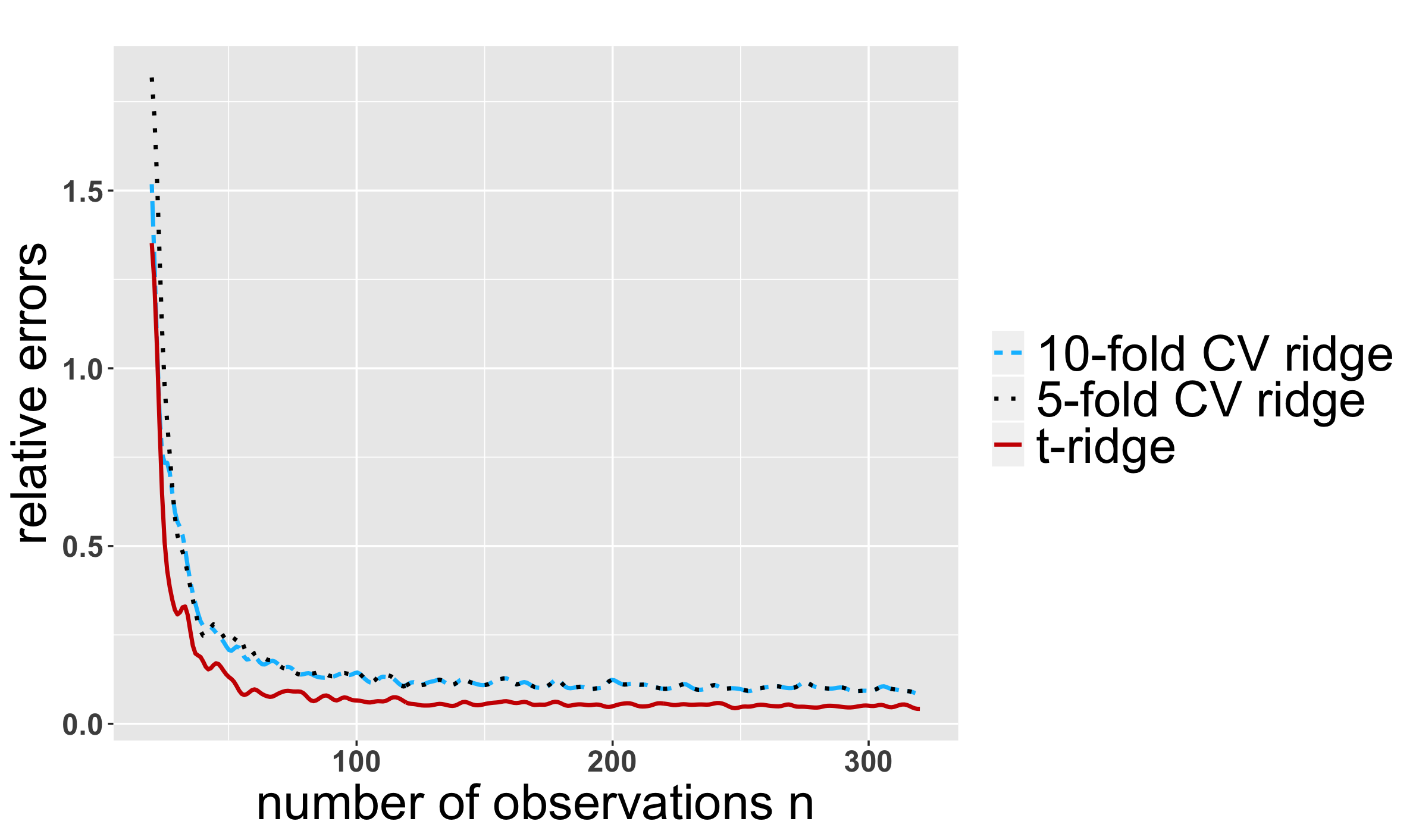} 
\caption[]{Relative errors of the t-ridge estimator as compared to the maximum likelihood estimator $\normtwo{\est_{\namemethod}-\est_{\operatorname{mle}}}/\normtwo{\est_{\operatorname{mle}}}$ for Gaussian data with  $\numvar=20$ and $k=0$.
The t-ridge estimator quickly approximates the maximum likelihood estimator when the sample size~$\samplesize$ increases.
}
\label{fig:gaussian}
\end{figure}

Figure~\ref{fig:gaussian} confirms that the t-ridge converges rapidly to the unregularized maximum likelihood estimator $\est_{\operatorname{mle}}$, which  minimizes $\objfcnTe\prt{\parameter|\outcome, \design}$ defined in~\eqref{eq:glmlog}, in the relative error $\normtwo{\est_{\namemethod}-\est_{\operatorname{mle}}}/\normtwo{\est_{\operatorname{mle}}}$ as $p=20$ is fixed and the number of observations~$n$ increases.
Similar observations can be made in the Poisson and Bernoulli cases.
These results suggest that t-ridge estimators can be applied without regard of the dimensionality of the problem.


\subsubsection{Poisson case}
\label{subsec:examplelpoisson}

We then generate Poisson data.
According to Table~\ref{table:glm},
the t-ridge  estimator~\eqref{eq:t-ridgeglm} is
\begin{equation*}
  \est_{\namemethod} \in\argmin\limits_{\parameter\in\Rp}\bcbbbb{\frac{-\sumin\prtb{\samyi\samxi\tp\parameter-\exp\prt{{\samxi\tp\parameter}}}}{\normtwo{\sumin\samxi\tp\prtb{\samyi-\exp\prt{{\samxi\tp\parameter}}}}}+\normtwo{\parameter}}.
\end{equation*}

\begin{table}[htb]
    \centering
    \caption{t-ridge outperforms $K$-fold cross-validated ridge ($K$-fold CV ridge) for $K \in \{5, 10\}$  in prediction on Poisson data with $k=0$.}
    \ra{1.2}
    \begin{tabular}{c c c c c c}
    \toprule
    \midrule
         \multirow{2}{*}{Relative prediction error} &
         \multirow{2}{*}{n} & 
         \multirow{2}{*}{p} & 
         \multicolumn{3}{c}{Mean of relative errors (sd)} \\ \cline{4-6}
         
         {} & {} & {} & t-ridge 
                      & 5-fold CV ridge
                      & 10-fold CV ridge \\
         \hline
         \multirow{3}{*}{$\frac{\normtwo{\design\prt{\est-\target}}}{\normtwo{\design\target}}$} &
              100 & 300 & 0.60 (0.07) & 0.83 (0.06) & 0.82 (0.06) \\ 
{} & 200 & 500 & 0.65 (0.06) & 0.79 (0.05) & 0.79 (0.05) \\ 
{} & 50 & 1000 & 0.84 (0.06) & 0.97 (0.03) & 0.97 (0.03) \\  
         \midrule
         \bottomrule
    \end{tabular}
    \label{table:poisson0}
\end{table}

Table \ref{table:poisson0} shows that the t-ridge estimator 
outperforms 5- and 10-fold cross-validated ridge across all settings.


\subsubsection{Bernoulli case}
\label{subsec:examplelbernoulli}

We finally generate Bernoulli data. 
According to Table~\ref{table:glm},
the t-ridge  estimator~\eqref{eq:t-ridgeglm} is
\begin{equation*}
 \est_{\namemethod} \in\argmin\limits_{\parameter\in\Rp}\bcbbbb{\frac{-\sumin\prtbb{\samyi\samxi\tp\parameter-\prtb{1 + \exp\prt{{\samxi\tp\parameter}}}}}{\normtwo{\sumin\samxi\tp\prtbb{\samyi-\frac{\exp\prt{{\samxi\tp\parameter}}}{1+\exp\prt{{\samxi\tp\parameter}}}}}}+\normtwo{\parameter}}.
\end{equation*}
\noindent 

\begin{table}[htb]
    \centering
    \caption{t-ridge rivals $K$-fold cross-validated ridge ($K$-fold CV ridge) for $K \in \{5, 10\}$  in prediction on Bernoulli data with $k=0$.}
    \ra{1.2}
    \begin{tabular}{c c c c c c}
    \toprule
    \midrule
         \multirow{2}{*}{Relative prediction error} &
         \multirow{2}{*}{n} & 
         \multirow{2}{*}{p} & 
         \multicolumn{3}{c}{Mean of relative errors (sd)} \\ \cline{4-6}
         
         {} & {} & {} & t-ridge 
                      & 5-fold CV ridge
                      & 10-fold CV ridge \\
         \hline
         \multirow{3}{*}{$\frac{\normtwo{\design\prt{\est-\target}}}{\normtwo{\design\target}}$} &
              100 & 300 & 0.87 (0.06) & 0.88 (0.07) & 0.88 (0.07) \\ 
{} & 200 & 500 & 0.86 (0.06) & 0.87 (0.07) & 0.86 (0.06) \\ 
{} & 50 & 1000 & 0.88 (0.08) & 0.91 (0.09) & 0.90 (0.10) \\  
         \midrule
         \bottomrule
    \end{tabular}
    \label{table:bernoulli0}
\end{table}

Table~\ref{table:bernoulli0} shows that the t-ridge estimators rival 5- and 10-fold cross-validated ridge across all settings.

Taken together,
the results in the Gaussian, Poisson, and Bernoulli case suggest that the t-ridge estimator is an alternative to standard pipelines for data of any dimension and type.

\clearpage 
\section{Discussion}
\label{sec:discussion}
We have shown that the calibration of the tuning parameter can be incorporated directly into the formulation of ridge estimators. 
Since our approach in Section~\ref{sec:methodology}, called t-ridge, requires essentially only that the data-fitting function is differentiable, 
it can be applied  to a  wide variety of ridge estimators.

As an example,
we have detailed   the t-ridge estimator in Section~\ref{sec:applications} for  generalized linear models,
and we complemented the theoretical insights of Section~\ref{sec:methodology}
to corroborate  the estimator's motivation further.
    We expect that these mathematical insights will also be of use for tuning parameter calibration beyond the ridge estimator.
We have also shown in Section~\ref{sec:numerical} that the t-ridge estimator can be implemented efficiently and that it can outperform standard pipelines empirically across different types of data and dimensions.

We finally expect that  tuning-free estimators such as trex and t-ridge can also be valuable for post-selection problems  \citep{ Taylor2014PostselectionAI,taylor_tibshirani_2017}
since the inclusion of  calibration schemes can be  difficult in such problems.

\section*{Acknowledgements}
We thank Jacob Bien, Klaus Holst, and Stefan Sperlich for their insightful comments.

\bibliographystyle{agsm}
\bibliography{Bibliography-MM-MC}


\begin{appendices}

\section{Proofs}
\label{subsec:proofs}

\subsection{Proof of Theorem~\ref{thm:ridge-edr}}
\label{subsec:ridge-edrProof}

\begin{proof}
We first show 1.\@, that is, for each ridge estimator $\esttheta_{\nameridge}[\ridgetuning]$ with $\ridgetuning \geq 0$, there always exists a $\tuning\ge0$ satisfying  $\tuning=2\ridgetuning\normtwo{\esttheta_{\nameridge}[\ridgetuning]}$ such that $\esttheta_{\nameedr}[\tuning]=\esttheta_{\nameridge}[\ridgetuning]$.

Using the Karush–Kuhn–Tucker~\citep{kuhn1951} conditions for both  \eqref{eq:generalizedRidge} and \eqref{eq:generalizedEDR}, we have 
\begin{equation*}
    -\derivative\prt{\esttheta_{\nameridge}[\ridgetuning]}+2\ridgetuning\esttheta_{\nameridge}[\ridgetuning]=\zero_{p} 
\end{equation*}
\noindent and 
\begin{equation*}
    -\derivative\prt{\esttheta_{\nameedr}[\tuning]}+\tuning\frac{\partiald\normtwo{\paratheta}}{\partiald\paratheta}\Big|_{\paratheta=\esttheta_{\nameedr}[\tuning]}=\zero_{p},
\end{equation*}
\noindent where the sub-differential of the $\ell_{2}$ norm with respect to $\paratheta$ is defined as 
\begin{equation*}
    \frac{\partiald\normtwo{\paratheta}}{\partiald\paratheta} :=
    \begin{cases}              \frac{\paratheta}{\normtwo{\paratheta}} \ &{\rm if} \ \paratheta \neq \zero_{p}\\
   \{\subdifferential \in \Rp : \normtwo{\subdifferential} \leq 1 \} \ &{\rm if} \ \paratheta = \zero_{p}.
    \end{cases}
\end{equation*} 
By rearrangement, we obtain 
\begin{equation}
\label{eq:kktridge}
    \derivative\prt{\esttheta_{\nameridge}[\ridgetuning]}=2\ridgetuning\esttheta_{\nameridge}[\ridgetuning],
\end{equation}
\noindent and 
\begin{equation}
\label{eq:kktedr}
    \derivative\prt{\esttheta_{\nameedr}[\tuning]}=\tuning\frac{\partiald\normtwo{\paratheta}}{\partiald\paratheta}\Big|_{\paratheta=\esttheta_{\nameedr}[\tuning]}.
\end{equation}
\noindent If $\esttheta_{\nameridge}[\ridgetuning]=\zero_{p}$, by~\eqref{eq:kktridge} we have 
\begin{equation}
\label{eq:zeroscore}
    \derivative\prt{\esttheta_{\nameridge}[\ridgetuning]}=\derivative\prt{\zero_\numvar}=\zero_\numvar.
\end{equation}
By setting $\tuning=0$, \eqref{eq:kktedr} yield 
\begin{equation*}
    \derivative\prt{\esttheta_{\nameedr}[0]}=\zero_\numvar.
\end{equation*}
\noindent The convexity of~\eqref{eq:generalizedRidge} and~\eqref{eq:generalizedEDR} imply that  $\esttheta_{\nameedr}[0]=\zero_\numvar=\esttheta_{\nameridge}[\ridgetuning]$. 
In addition, taking $\ell_2$ norm on both sides of equations~\eqref{eq:kktridge} and~\eqref{eq:kktedr} yields 
\begin{equation*}
    \normtwo{\derivative\prt{\esttheta_{\nameridge}[\ridgetuning]}}=0=\normtwo{\derivative\prt{\esttheta_{\nameedr}[0]}}.
\end{equation*}
\noindent If $\esttheta_{\nameridge}[\ridgetuning]\neq\zero_\numvar$, letting $\tuning=2\ridgetuning\normtwo{\esttheta_{\nameridge}[\ridgetuning]}$ and by ~\eqref{eq:kktridge}, we have 
\begin{align*}
    \derivative\prt{\esttheta_{\nameridge}[\ridgetuning]}&=2\ridgetuning\normtwo{\esttheta_{\nameridge}[\ridgetuning]}\frac{\esttheta_{\nameridge}[\ridgetuning]}{\normtwo{\esttheta_{\nameridge}[\ridgetuning]}}\\
    &=\tuning\frac{\esttheta_{\nameridge}[\ridgetuning]}{\normtwo{\esttheta_{\nameridge}[\ridgetuning]}}.
\end{align*}
\noindent Comparing the equation above with~\eqref{eq:kktedr}, we observe that  $\esttheta_{\nameedr}[\tuning]=\esttheta_{\nameridge}[\ridgetuning]$ can be a solution of~\eqref{eq:kktedr}. 
Furthermore, by taking $\ell_2$ norm on both sides of equations~\eqref{eq:kktridge} and~\eqref{eq:kktedr}, we have 
\begin{equation*}
    \normtwo{\derivative\prt{\esttheta_{\nameridge}[\ridgetuning]}}=\tuning=\normtwo{\derivative\prt{\esttheta_{\nameedr}[\tuning]}}
\end{equation*}
\noindent as desired.

Secondly, we prove 2.\@, that is, for each edr estimator $\esttheta_{\nameedr}[\tuning]$ with $\tuning \geq 0$, there always exists a $\ridgetuning\ge0$ satisfying  $\ridgetuning=\tuning/(2\normtwo{\esttheta_{\nameedr}[\tuning]})$ such that $\esttheta_{\nameridge}[\ridgetuning]=\esttheta_{\nameedr}[\tuning]$.

If $\esttheta_{\nameedr}[\tuning]=\zero_{p}$ and $\tuning \neq 0$, we can find $\ridgetuning=\tuning/(2\normtwo{\esttheta_{\nameedr}[\tuning]})=\infty$ such that $\esttheta_{\nameridge}[\infty]=\zero_{\numvar}$, which is shown in the following. 
Let $\ridgetuning=\infty$, 
then by the definition in \eqref{eq:generalizedRidge}, we have for any $\paratheta \in \Rp$ 
\begin{equation*}
    \objfcnTe\prt{\esttheta_{\nameridge}[\infty]|\data}+\infty\cdot\normtwo{\esttheta_{\nameridge}[\infty]}^2 
    \leq \objfcnTe\prt{\paratheta|\data}+\infty\cdot\normtwo{\paratheta}^2.
\end{equation*}
However, we observe that 
\begin{align*}
    \objfcnTe\prt{\zero_{\numvar}|\data}+\infty\cdot\normtwo{\zero_{\numvar}}^2 
    &=\objfcnTe\prt{\zero_{\numvar}|\data} + \tuning\normtwo{\zero_{\numvar}}\\
    &\leq \objfcnTe\prt{\paratheta|\data}+\tuning\normtwo{\paratheta} \\
    &\leq \objfcnTe\prt{\paratheta|\data}+\infty\cdot\normtwo{\paratheta}^2
\end{align*}
\noindent holds for any vector $\paratheta \in \Rp$.
By the convexity of the data-fitting function $\objfcnTe$, we know that $\esttheta_{\nameridge}[\ridgetuning]$ is unique and hence, $\esttheta_{\nameridge}[\ridgetuning]=\zero_{\numvar}$.

If $\esttheta_{\nameedr}[0]=\zero_{p}$ and $\tuning=0$, then we have
\begin{align*}
    \objfcnTe\prt{\zero_{\numvar}|\data}+0\cdot\normtwo{\zero_{\numvar}}^2 
    &=\objfcnTe\prt{\zero_{\numvar}|\data} + 0\cdot\normtwo{\zero_{\numvar}}\\
    &\leq \objfcnTe\prt{\paratheta|\data}+0\cdot\normtwo{\paratheta} \\
    &= \objfcnTe\prt{\paratheta|\data}+0\cdot\normtwo{\paratheta}^2
\end{align*}
\noindent Hence, $\esttheta_{\nameridge}[0]=\zero_{\numvar}$.

If $\esttheta_{\nameedr}[\tuning]\neq\zero_\numvar$, letting $\ridgetuning=\tuning/(2\normtwo{\esttheta_{\nameedr}[\tuning]})$ and by ~\eqref{eq:kktedr}, we have 
\begin{align*}
    \derivative\prt{\esttheta_{\nameedr}[\tuning]}&=2\frac{\tuning}{2\normtwo{\esttheta_{\nameedr}[\tuning]}}\esttheta_{\nameedr}[\tuning]\\
    &=2\ridgetuning\esttheta_{\nameedr}[\tuning].
\end{align*}
\noindent Comparing the equation above with~\eqref{eq:kktridge}, we observe that $\esttheta_{\nameridge}[\ridgetuning]=\esttheta_{\nameedr}[\tuning]$ can be a solution of~\eqref{eq:kktridge}. 
Again, by taking $\ell_2$ norm on both sides of equations~\eqref{eq:kktridge} and~\eqref{eq:kktedr}, we also have 
\begin{equation*}
    \normtwo{\derivative\prt{\esttheta_{\nameridge}[\ridgetuning]}}=\tuning=\normtwo{\derivative\prt{\esttheta_{\nameedr}[\tuning]}}
\end{equation*}
\noindent as desired.
\end{proof}


\subsection{Proof of Theorem~\ref{thm:edr-tridge}}
\label{subsec:edr-tridgeProof}
\begin{proof}
We prove the theorem by two steps:
\begin{enumerate}
    \item
    $\esttheta_{\namemethod}=\esttheta_{\nameedr}[\tuning]$ where $\tuning=\normtwo{\derivative\prt{\esttheta_{\namemethod}}}$;
    \item 
    $\esttheta_{\nameedr}[\tuning]=\esttheta_{\nameridge}[\ridgetuning]$ where $\ridgetuning=\tuning/(2\normtwo{\esttheta_{\nameedr}[\tuning]})$. 
\end{enumerate}
Firstly, we prove 1.\@. 
By the definition of t-ridge estimator in \eqref{eq:generalizedt-ridge}, we have
\begin{align*}
    \frac{\objfcnTe\prt{\esttheta_{\namemethod}|\data}}{\normtwo{\derivative\prt{\esttheta_{\namemethod}}}}+\normtwo{\esttheta_{\namemethod}}\le\frac{\objfcnTe\prt{\esttheta_{\nameedr}[\tuning]|\data}}{\normtwo{\derivative\prt{\esttheta_{\nameedr}[\tuning]}}}+\normtwo{\esttheta_{\nameedr}[\tuning]}.
\end{align*}
Notice that $\tuning=\normtwo{\derivative\prt{\esttheta_{\namemethod}}}\neq0$. 
By Theorem~\ref{thm:ridge-edr}, we have
\begin{equation*}
    \normtwo{\derivative\prt{\esttheta_{\namemethod}}}=\tuning=\normtwo{\derivative\prt{\esttheta_{\nameedr}[\tuning]}}.
\end{equation*}
Using this relation and multiplying $\tuning$ on the both sides of above inequality yields 
\begin{equation*}
    \objfcnTe\prt{\esttheta_{\namemethod}|\data}+\tuning\normtwo{\esttheta_{\namemethod}}\le\objfcnTe\prt{\esttheta_{\nameedr}[\tuning]|\data}+\tuning\normtwo{\esttheta_{\nameedr}[\tuning]}.
\end{equation*}
On the other hand, by the definition of edr estimator, we have 
\begin{equation*}
    \objfcnTe\prt{\esttheta_{\nameedr}[\tuning]|\data}+\tuning\normtwo{\esttheta_{\nameedr}[\tuning]}\le\objfcnTe\prt{\esttheta_{\namemethod}|\data}+\tuning\normtwo{\esttheta_{\namemethod}}.
\end{equation*}
Combining the two inequalities above yields 
\begin{equation*}
    \objfcnTe\prt{\esttheta_{\namemethod}|\data}+\tuning\normtwo{\esttheta_{\namemethod}}=\objfcnTe\prt{\esttheta_{\nameedr}[\tuning]|\data}+\tuning\normtwo{\esttheta_{\nameedr}[\tuning]}.
\end{equation*}
Since we assume the data-fitting function $\objfcnTe$ is convex, the objective function of edr method is also convex, which means edr has a unique global minimum. 
Hence, we have $\esttheta_{\namemethod}=\esttheta_{\nameedr}[\tuning]$, which proved 1.\@.

Since the second equality $\esttheta_{\nameedr}[\tuning]=\esttheta_{\nameridge}[\ridgetuning]$ can be obtained directly by Theorem~\ref{thm:ridge-edr}, we finish the prove.
\end{proof}


\subsection{Proof of Theorem~\ref{thm:uniqueness}}
\label{subsec:uniqueness}
\begin{proof}
We prove this theorem by contradiction. 
Suppose there are two t-ridge estimators $\esttheta_{\namemethod}'$ and $\esttheta_{\namemethod}''$ such that $\esttheta_{\namemethod}'\neq\esttheta_{\namemethod}''$. 
Let $\tuning'=\normtwo{\derivative\prt{\esttheta_{\namemethod}'}}$ and $\tuning''=\normtwo{\derivative\prt{\esttheta_{\namemethod}''}}$.
Note that the definition of t-ridge object function implies that the $\ell_{2}$ norm of the score function with respect to the t-ridge estimator is non-zero.
Hence, $\tuning'=\tuning''\neq 0$. 
By Theorem~\ref{thm:edr-tridge}, we have 
\begin{equation*}
    \esttheta_{\namemethod}'=\esttheta_{\nameedr}[\tuning']=\esttheta_{\nameedr}[\tuning'']=\esttheta_{\namemethod}'',
\end{equation*}
which produces a contradiction with $\esttheta_{\namemethod}'\neq\esttheta_{\namemethod}''$. 

So, we aim to show in the following that if both $\esttheta_{\namemethod}'$ and $\esttheta_{\namemethod}''$ minimize the objective function of t-ridge, then $\tuning'=\tuning''$.
By Theorem~\ref{thm:edr-tridge}, we know that $\esttheta_{\namemethod}'=\esttheta_{\nameedr}[\tuning']$ and $\esttheta_{\namemethod}''=\esttheta_{\nameedr}[\tuning'']$. 
Since $\esttheta_{\namemethod}'$ and $\esttheta_{\namemethod}''$ are minimums of t-ridge, we have \begin{align*}
    \frac{\objfcnTe\prt{\esttheta_{\nameedr}[\tuning']|\data}}{\tuning'}+\normtwo{\esttheta_{\nameedr}[\tuning']}&=\frac{\objfcnTe\prt{\esttheta_{\nameedr}[\tuning'']|\data}}{\tuning''}+\normtwo{\esttheta_{\nameedr}[\tuning'']}\\
    &=\frac{1}{\tuning''}\prtb{\objfcnTe\prt{\esttheta_{\nameedr}[\tuning'']|\data}+\tuning''\normtwo{\esttheta_{\nameedr}[\tuning'']}}\\
    &\le\frac{1}{\tuning''}\prtb{\objfcnTe\prt{\esttheta_{\nameedr}[\tuning']|\data}+\tuning''\normtwo{\esttheta_{\nameedr}[\tuning']}}\\
    &=\frac{\objfcnTe\prt{\esttheta_{\nameedr}[\tuning']|\data}}{\tuning''}+\normtwo{\esttheta_{\nameedr}[\tuning']}.
\end{align*}
This implies 
\begin{equation*}
    \frac{\objfcnTe\prt{\esttheta_{\nameedr}[\tuning']|\data}}{\tuning'}\le\frac{\objfcnTe\prt{\esttheta_{\nameedr}[\tuning']|\data}}{\tuning''}.
\end{equation*}
Similarly, we can obtain 
\begin{equation*}
    \frac{\objfcnTe\prt{\esttheta_{\nameedr}[\tuning'']|\data}}{\tuning''}\le\frac{\objfcnTe\prt{\esttheta_{\nameedr}[\tuning'']|\data}}{\tuning'}.
\end{equation*}
By the assumption that $\objfcnTe\prt{\paratheta|\data}>0$ for all $\paratheta \in \Rp$, we have $\objfcnTe\prt{\esttheta_{\nameedr}[\tuning']|\data}>0$ and $\objfcnTe\prt{\esttheta_{\nameedr}[\tuning'']|\data}>0$. Hence, the two inequalities above yield $\tuning'=\tuning''$ and we get the desired result.
\end{proof}

\subsection{Proof of Theorem~\ref{thm:edrBound}}
\label{subsec:edrboundProof}

\begin{proof}
By the definition of~\eqref{eq:glmEdr}, we have
\begin{equation*}
    -\sumin\prtb{\samyi\samxi\tp\est_{\nameedr}[\tuning]-\fcnb{\samxi\tp\est_{\nameedr}[\tuning]}}+\tuning\normtwo{\est_{\nameedr}[\tuning]}
    \le -\sumin\prtb{\samyi\samxi\tp\target-\fcnb{\samxi\tp\target}}+\tuning\normtwo{\target}.
\end{equation*}
\noindent By arranging, we have
\begin{equation*}
    \sumin\prt{\fcnb{\samxi\tp\est_{\nameedr}[\tuning]}-\fcnb{\samxi\tp\target}}
    \le\sumin\samyi\samxi\tp\prt{\est_{\nameedr}[\tuning]-\target]}+\tuning\prt{\normtwo{\target}-\normtwo{\est_{\nameedr}[\tuning]}}.
\end{equation*}
\noindent The margin condition on $\fcnnb$ yields
\begin{equation*}
    \fcnb{\samxi\tp\est_{\nameedr}[\tuning]}-\fcnb{\samxi\tp\target}
    \ge\fcnnb'\prt{\samxi\tp\target}\prt{\samxi\tp\est_{\nameedr}[\tuning]-\samxi\tp\target}
    +\frac{1}{\margincondC^2}\prt{\samxi\tp\est_{\nameedr}[\tuning]-\samxi\tp\target}^2, 
\end{equation*}
\noindent where $\margincondC=max_{i \in \{1, \cdots, \samplesize\}}\{C_{i}\}$.
Notice $\fcnnb'\prt{\samxi\tp\target}=\linkfcnn\inv\prt{\samxi\tp\target}$ and $\samyi = \linkfcnn\inv\prt{\samxi\tp\target} + \varei$, we can obtain
\begin{equation*}
    \frac{1}{\margincondC^2}\sumin\prt{\samxi\tp\prt{\est_{\nameedr}[\tuning]-\target}}^2
    \le\sumin\varei\prt{\samxi\tp\prt{\est_{\nameedr}[\tuning]-\target]}}+\tuning\prt{\normtwo{\target}-\normtwo{\est_{\nameedr}[\tuning]}}.
\end{equation*}
\noindent We write it as the matrix form
\begin{equation*}
    \frac{1}{\margincondC^2}\normtwos{\design\prt{\est_{\nameedr}[\tuning]-\target}}
    \le\langle\design\tp\vare,\est_{\nameedr}[\tuning]-\target]\rangle+\tuning\prt{\normtwo{\target}-\normtwo{\est_{\nameedr}[\tuning]}}.
\end{equation*}
Using the H\"older's inequality on the first term of right hand side, we get
\begin{equation*}
    \frac{1}{\margincondC^2}\normtwos{\design\prt{\est_{\nameedr}[\tuning]-\target}}
    \le\normtwo{\design\tp\vare}\normtwo{\est_{\nameedr}[\tuning]-\target]}+\tuning\prt{\normtwo{\target}-\normtwo{\est_{\nameedr}[\tuning]}}.
\end{equation*}
\noindent By assumption $\tuning\ge\tuning^*$ and triangle inequality, we obtain
\begin{align*}
    \normtwos{\design\prt{\est_{\nameedr}[\tuning]-\target}}
    &\le\tuning\margincondC^2\prtbb{\normtwo{\est_{\nameedr}[\tuning]-\target]}+\normtwo{\target}-\normtwo{\est_{\nameedr}[\tuning]}}\\
    &\le2\tuning\margincondC^2\normtwo{\target}
\end{align*}
as desired.
\end{proof}

\subsection{Proof of Theorem~\ref{thm:t-ridge}}
\label{subsec:tridgeProof}

\begin{proof}
By Theorem \ref{thm:edr-tridge}, we know that t-ridge is on the ridge path and hence, it is also on the edr path by Theorem \ref{thm:ridge-edr} such that $\est_{\namemethod}=\est_{\nameedr}[\esttuning]$.
According to the definition of $\est_{\nameedr}$, we have
\begin{equation*}
    -\sumin\prtb{\samyi\samxi\tp\est_{\nameedr}[\esttuning]-\fcnb{\samxi\tp\est_{\nameedr}[\esttuning]}}+\tuning^*\normtwo{\est_{\nameedr}[\esttuning]}
    \le-\sumin\prtb{\samyi\samxi\tp\target-\fcnb{\samxi\tp\target}}+\tuning^*\normtwo{\target}.
\end{equation*}
\noindent Following the proof of Theorem \ref{thm:edrBound} (the penultimate inequality), we have
\begin{align*}
    \frac{1}{\margincondC^2}\normtwos{\design\prt{\est_{\nameedr}[\esttuning]-\target}}
    &\le\tuning^{*}\normtwo{\est_{\nameedr}[\esttuning]-\target}+\esttuning\prt{\normtwo{\target}-\normtwo{\est_{\nameedr}[\esttuning]}}\\
    &\le 2\max\{\tuning^{*},\esttuning\}\normtwo{\target}.
\end{align*}
\noindent Multiplying $\margincondC^2$ on both sides yields the desired result.

\end{proof}

\subsection{Proof of Lemma~\ref{lem:checktridge}}
\label{subsec:checktridge}

\begin{proof}
According to the definition of t-ridge in \eqref{eq:t-ridgeglm}, we have 
\begin{equation*}
    \frac{\objfcnTe\prt{\est_{\namemethod}|\outcome, \design}}{\esttuning} + \normtwo{\est_{\namemethod}}
    \leq 
    \frac{\objfcnTe\prt{\est_{\nameedr}[\tuning^{*}]|\outcome, \design}}{\tuning^{*}} + \normtwo{\est_{\nameedr}[\tuning^{*}]}.
\end{equation*}
\noindent If we multiply $\tuning^{*}$ on both sides of the above inequality, we obtain
\begin{equation*}
    \frac{\tuning^{*}}{\esttuning}\objfcnTe\prt{\est_{\namemethod}|\outcome, \design} + \tuning^{*}\normtwo{\est_{\namemethod}}
    \leq
    \objfcnTe\prt{\est_{\nameedr}[\tuning^{*}]|\outcome, \design} + \tuning^{*}\normtwo{\est_{\nameedr}[\tuning^{*}]}.
\end{equation*}
\noindent Hence, 
\begin{equation*}
\label{eq:checktridge}
    \prtbb{\frac{\tuning^{*}}{\esttuning}- 1}\objfcnTe\prt{\est_{\namemethod}|\outcome, \design} +
    \objfcnTe\prt{\est_{\namemethod}|\outcome, \design} +
    \tuning^{*}\normtwo{\est_{\namemethod}}
    \leq 
    \objfcnTe\prt{\est_{\nameedr}[\tuning^{*}]|\outcome, \design} + \tuning^{*}\normtwo{\est_{\nameedr}[\tuning^{*}]}.
\end{equation*}
\noindent By the definition of edr in \eqref{eq:glmEdr}, we know that 
\begin{equation*}
   \objfcnTe\prt{\est_{\nameedr}[\tuning^{*}]|\outcome, \design} + \tuning^{*}\normtwo{\est_{\nameedr}[\tuning^{*}]}
   \leq
    \objfcnTe\prt{\est_{\namemethod}|\outcome, \design} +
    \tuning^{*}\normtwo{\est_{\namemethod}}.
\end{equation*}
\noindent Combining this with the previous inequality yields 
\begin{equation*}
    \prtbb{\frac{\tuning^{*}}{\esttuning}- 1}\objfcnTe\prt{\est_{\namemethod}|\outcome, \design} \le 0.
\end{equation*}
If $\objfcnTe\prt{\est_{\namemethod}|\outcome, \design} > 0$, this implies that $\esttuning \ge \tuning^{*}$. 
If $\objfcnTe\prt{\est_{\namemethod}|\outcome, \design} < 0$, this implies that $\esttuning \le \tuning^{*}$. 
We finish the proof.

\end{proof}


\section{Additional Simulations}
\label{sec:addition}
Tables~\ref{table:gaussian}--\ref{table:bernoulli} give the results for the remaining settings described in Section~\ref{sec:numerical}.
These results further corroborate our conclusion that the t-ridge estimator is a contender across dimensions and data types.
\begin{table}[h]
    \centering
    \caption{t-ridge outperforms $K$-fold cross-validated ridge ($K$-fold CV ridge) for $K \in \{5, 10\}$  in prediction on  Gaussian data with $k \in \{0.2, 0.4\}$.}
    \ra{1.2}
    \begin{tabular}{c c c c c c c}
    \toprule
    \midrule
         \multirow{2}{*}{Relative prediction error} &
         \multirow{2}{*}{n} & 
         \multirow{2}{*}{p} & 
         \multirow{2}{*}{k} &
         \multicolumn{3}{c}{Mean of relative errors (sd)} \\ \cline{5-7}
         
         {} & {} & {} & {} 
                      & t-ridge 
                      & 5-fold CV ridge
                      & 10-fold CV ridge \\
         \hline
         \multirow{6}{*}{$\frac{\normtwo{\design\prt{\est-\target}}}{\normtwo{\design\target}}$} &
                 100 & 300 & 0.2 & 0.35 (0.03) & 0.54 (0.06) & 0.53 (0.05) \\ 
{} & 100 & 300 & 0.4 & 0.36 (0.03) & 0.53 (0.03) & 0.53 (0.03) \\ 
{} & 200 & 500 & 0.2 & 0.36 (0.02) & 0.50 (0.02) & 0.50 (0.02) \\ 
{} & 200 & 500 & 0.4 & 0.38 (0.02) & 0.50 (0.02) & 0.50 (0.02) \\ 
{} & 50 & 1000 & 0.2 & 0.32 (0.03) & 0.52 (0.26) & 0.50 (0.25) \\ 
{} & 50 & 1000 & 0.4 & 0.32 (0.04) & 0.55 (0.25) & 0.54 (0.26) \\  
         \midrule
         \bottomrule
    \end{tabular}
    \label{table:gaussian}
\end{table}

\begin{table}
    \centering
    \caption{t-ridge outperforms $K$-fold cross-validated ridge ($K$-fold CV ridge) for $K \in \{5, 10\}$  in prediction on Poisson data with $k \in \{0.2, 0.4\}$.}
    \ra{1.2}
    \begin{tabular}{c c c c c c c}
    \toprule
    \midrule
         \multirow{2}{*}{Relative prediction error} &
         \multirow{2}{*}{n} & 
         \multirow{2}{*}{p} &
         \multirow{2}{*}{k} &
         \multicolumn{3}{c}{Mean of relative errors (sd)} \\ \cline{5-7}
         
         {} & {} & {} & {} 
                      & t-ridge 
                      & 5-fold CV ridge
                      & 10-fold CV ridge \\
         \hline
         \multirow{6}{*}{$\frac{\normtwo{\design\prt{\est-\target}}}{\normtwo{\design\target}}$} &
                 100 & 300 & 0.2 & 0.63 (0.08) & 0.82 (0.05) & 0.82 (0.06) \\ 
{} & 100 & 300 & 0.4 & 0.64 (0.08) & 0.82 (0.05) & 0.81 (0.05) \\ 
{} & 200 & 500 & 0.2 & 0.66 (0.07) & 0.79 (0.05) & 0.78 (0.05) \\ 
{} & 200 & 500 & 0.4 & 0.66 (0.07) & 0.78 (0.04) & 0.77 (0.04) \\ 
{} & 50 & 1000 & 0.2 & 0.86 (0.05) & 0.96 (0.03) & 0.96 (0.03) \\ 
{} & 50 & 1000 & 0.4 & 0.86 (0.04) & 0.96 (0.03) & 0.96 (0.03) \\    
         \midrule
         \bottomrule
    \end{tabular}
    \label{table:poisson}
\end{table}

\begin{table}
    \centering
    \caption{t-ridge rivals $K$-fold cross-validated ridge ($K$-fold CV ridge) for $K \in \{5, 10\}$  in prediction on Bernoulli data with $k \in \{0.2, 0.4\}$.}
    \ra{1.2}
    \begin{tabular}{c c c c c c c}
    \toprule
    \midrule
         \multirow{2}{*}{Relative prediction  error} &
         \multirow{2}{*}{n} & 
         \multirow{2}{*}{p} & 
         \multirow{2}{*}{k} & 
         \multicolumn{3}{c}{Mean of relative errors (sd)} \\ \cline{5-7}
         
         {} & {} & {} & {} 
                      & t-ridge 
                      & 5-fold CV ridge
                      & 10-fold CV ridge \\
         \hline
         \multirow{6}{*}{$\frac{\normtwo{\design\prt{\est-\target}}}{\normtwo{\design\target}}$} &
                100 & 300 & 0.2 & 0.86 (0.06) & 0.88 (0.07) & 0.88 (0.08) \\ 
{} & 100 & 300 & 0.4 & 0.85 (0.06) & 0.86 (0.08) & 0.85 (0.08) \\ 
{} & 200 & 500 & 0.2 & 0.86 (0.05) & 0.87 (0.06) & 0.86 (0.06) \\ 
{} & 200 & 500 & 0.4 & 0.83 (0.05) & 0.84 (0.06) & 0.84 (0.06) \\ 
{} & 50 & 1000 & 0.2 & 0.87 (0.08) & 0.90 (0.09) & 0.89 (0.09) \\ 
{} & 50 & 1000 & 0.4 & 0.86 (0.08) & 0.88 (0.09) & 0.88 (0.09) \\   
         \midrule
         \bottomrule
    \end{tabular}
    \label{table:bernoulli}
\end{table}


\end{appendices}

\end{document}